\documentstyle[preprint,aps,epsf]{revtex}
\preprint{SPhT/95-149}

\begin{document}
\draft
\title{GENERALIZED HELIMAGNETS BETWEEN TWO AND FOUR DIMENSIONS}
\author{Fran\c cois David$^*$ and Thierry Jolic\oe ur\thanks{CNRS}}
\address{Service de Physique Th\'eorique, CE Saclay,\\ F-91191 Gif-sur-Yvette
CEDEX, France}
\date{\today}
\maketitle
\begin{abstract}
We study the phase transitions of N-components generalized helimagnets that
obey the symmetry breaking pattern
${\rm O}(N)\times {\rm O}(N-1)\rightarrow {\rm O}(N-1)_{\rm diag}$. In the
neighborhood of two dimensions, a $D=2+\epsilon$ renormalization group study
reveals a rich fixed point structure as well as a nematic-like phase with
partial spin ordering. In the physical case
${\rm O}(3)\times {\rm O}(2)\rightarrow {\rm O}(2)_{\rm diag}$, relevant to
real magnets with noncollinear ordering, we show that this implies an XY-like
transition between the ordered phase and the nematic-like phase. A non-Abelian
mean-field calculation qualitatively valid above four dimensions is shown to
lead to the same picture but then the principal chiral fixed point, which had
been proposed earlier as the relevant fixed point for D=3 helimagnets, plays no
role due to the appearance of a first-order line.
\end{abstract}

Several physical systems display a peculiar critical behaviour associated with
helical or noncollinear ordering. These include helimagnets, the dipole-locked
phase of superfluid Helium as well as Josephson junction arrays in a transverse
magnetic field\cite{plum}. A prototypical example is the stacked triangular
lattice antiferromagnet (STA) whose low-temperature ordered phase breaks
completely the rotation symmetry group. When generalized to N-component spins,
the Landau-Ginzburg free energy for this system contains two quartic
invariants: \iftrue
\begin{eqnarray}
H=&&{1 \over 2}\left( {\left( {\nabla \vec \varphi _1} \right)^2+\left( {\nabla
\vec \varphi _2} \right)^2} \right)+r\left( {\vec \varphi _1^2+\vec \varphi
_2^2} \right)
\nonumber\\
&& +u\left( {\vec \varphi _1^2+\vec \varphi _2^2} \right)^2+v\left( {\left(
{\vec \varphi _1\cdot\vec \varphi _2} \right)^2-\vec \varphi _1^2\,\vec \varphi
_2^2} \right). \label{LG}
\end{eqnarray}
\else
\begin{equation}
H={1 \over 2}\left( {\left( {\nabla \vec \varphi _1} \right)^2+\left( {\nabla
\vec \varphi _2} \right)^2} \right)+r\left( {\vec \varphi _1^2+\vec \varphi
_2^2} \right) +u\left( {\vec \varphi _1^2+\vec \varphi _2^2} \right)^2+v\left(
{\left( {\vec \varphi _1\cdot\vec \varphi _2} \right)^2-\vec \varphi _1^2\vec
\varphi _2^2} \right). \label{LG}
\end{equation}
\fi
When $u, v >0$, this free energy describes a second-order phase transition
between a high-temperature phase which is ${\rm O}(N)\times {\rm
O}(2)$-symmetric
and a low-temperature phase with lower symmetry ${\rm O}(N-2)\times {\rm
O}(2)_{\rm diag}$.
This theory has been studied in the neighborhood of four dimensions by the
standard $\epsilon$-expansion\cite{garel,kawa1}. The main point of these
studies is that
when the number of components is large
enough there is a fixed point with $u^*, v^* \neq 0$ that describes a critical
behaviour different from the well-studied behaviour of the N-vector model.
In the neighborhood of the upper critical dimension, D=4, there is a dividing
line $N_c (D) = 21.8-23.4\epsilon +O(\epsilon^2)$ in the (N, D)-plane above
which there is a second-order phase transition and below which there is a
fluctuation-induced first-order transition (no stable fixed point).
This is similar to the case of the normal-superconducting phase
transition\cite{halubma}.
The fate of the line $N_c(D)$ is not known outside the $\epsilon$-expansion.
Numerical studies on the STA lattice (thus in D=3) have shown that the ordering
transition is second-order with exponents that do not belong to the ${\rm
O}(N)$ Wilson-Fisher universality classes\cite{kawa2,bata,diep1}, $\nu
=0.585(9)$, $\gamma /\nu =2.011(14)$\cite{bata}. These exponents are also
observed in the body-centered tetragonal antiferromagnet\cite{diep2},
suggestive of a new universality class, usually called chiral universality
class. It has been suggested\cite{kawa1,kawa2,Jojo} that these exponents are
ruled by the nontrivial fixed point which exists for $N>N_c(D)$.

The study of these phenomena has been also performed from the low-temperature
expansion.
A non-linear sigma model that captures the Goldstone modes of the theory has
been constructed by Dombre and Read\cite{dombre}. It involves an order
parameter which is a rotation matrix instead of a vector as in usual magnetic
systems.
This is due to the fact that the rotation group is fully broken in the
low-temperature phase of noncollinear magnets. The symmetry breaking pattern is
${\rm SO}(3)\times {\rm SO}(2)\rightarrow {\rm SO}(2)_{\rm diag}$ where the
internal ${\rm SO}(2)$ rotation acts upon the 1,2 indices of the fields of
Eq.(\ref{LG}).
We can parametrize the rotation matrix of the non-linear sigma model by three
orthogonal unit vectors
${\vec \phi}_1 , {\vec \phi}_2 ,{\vec \phi}_3$ and the Euclidean action can be
then written as: \iffalse
\begin{eqnarray}
\end{eqnarray}
\else
\begin{equation}
S=\int\! d^D x\, {1\over 2g_1}\left(( \nabla {\vec \phi}_1)^2 +( \nabla {\vec
\phi}_2)^2\right) +{1\over 2g_2}( \nabla {\vec \phi}_3)^2 .
\label{sigma3}
\end{equation}
\fi
The STA corresponds to a bare value $1/g_2 =0$. The renormalization group flow
of the two couplings $g_1 , g_2$ has been studied\cite{aza} in a $D=2+\epsilon$
expansion. It has been noted that a remarkable symmetry enhancement happens on
the line $g_1 =g_2$.
In this case the global symmetry of Eq.\ (\ref{sigma3}) is ${\rm SO}(3)\times
{\rm SO}(3)$ broken down to the diagonal subgroup ${\rm SO}(3)$.
This is the so-called principal chiral model which is renormalizable and thus
the peculiar line $g_1 =g_2$ is stable under the RG flow. Since ${\rm
SO}(3)\times {\rm SO}(3)\equiv {\rm SO}(4)$, this line is described by the
usual ${\rm SO}(4)$ sigma model, at least within the $D=2+\epsilon$ expansion.
Since this fixed point is stable and has an exponent $\nu\approx 0.74$ there is
a clear conflict with numerical and experimental findings in $D=3$.

In this Letter we shed new light on this problem by studying  a
new generalization to N components of this sigma model  as well as by a
non-Abelian mean-field calculation.
We show that the sigma model has a much richer
structure than previously expected and we propose a scenario in which the
principal chiral fixed point with ${\rm SO}(4)$ symmetry plays no role in $D=3$
due to the appearance
of a first-order transition.

In the two-dimensional case, the perturbative beta functions\cite{Friedan,aza}
indicate a flow which is infrared unstable away from the origin $g_1 =g_2 =0$
but the whole line $g_2 =0$ is a line of unstable fixed points as can read from
the RG formulas.
This fact has a simple interpretation: when $g_2 \rightarrow 0$ the vector
${\vec \phi}_3$ becomes frozen and the remaining degrees of freedom are simply
the rotations of the vectors ${\vec \phi}_1 ,{\vec \phi}_2$ in the plane
orthogonal to ${\vec \phi}_3$.
Since ${\vec \phi}_1$ and ${\vec \phi}_2$ are themselves orthogonal, we are
left with an XY model whose coupling is $g_1$. The corresponding beta function
is zero in perturbation theory. However it is important to note that in $D=2$
this XY model will undergo the Kosterlitz-Thouless (KT) transition for a {\it
finite} value of $g_1$. This phenomenon is not seen in perturbation theory and
is not found in the perturbative beta functions.
We expect that the line of fixed points $g_2 =0$ will end at some critical
value.
Above this value, the flow will go to the high-temperature fixed point in the
disordered phase of the XY model.
This transition is the unbinding transition of the vortices of the XY model. If
we unfreeze the vector ${\vec\phi}_3$ by setting $g_2\neq 0$, then topological
defects survive because, as noted by Kawamura and Miyashita\cite{KM}, $\Pi_1
({\rm SO}(3)) = Z_2$. All XY vortices with even winding number become
topologically trivial while those with odd winding number become all equivalent
and nontrivial.

To complete the phase diagram in the ($g_1 -g_2$)-plane, we note that for $g_1
=\infty$ the model (\ref{sigma3}) becomes the ${\rm O}(3)$ sigma model which
has no transition in D=2: the coupling $g_2$ flows continuously to the
high-temperature fixed point. If we assume that there are no other fixed
points, we are led to propose the phase diagram shown in Fig. 1. This means
that for nonzero values of the couplings the model is always disordered and
there is no phase transition at nonzero temperature as expected for a system
with non-abelian symmetry.

We note that the KT point on the $g_1$ axis cannot be studied by extending the
model simply to $N$ components.
The low-temperature limit of the theory (1) is an ${\rm O}(N)\times {\rm O}(2)$
sigma model\cite{nuke} that has also two coupling constants. The corresponding
phase diagram is topologically similar to Fig.~1: the KT point remains at
finite coupling unreachable by perturbation theory.

To clarify the situation, we will study another generalization of the ${\rm
O}(3)\times {\rm O}(2)$
model which involves $N$ orthogonal unit vectors with $N$ components
${\vec\phi_1},\dots,{\vec\phi_N}$
and the action is:
\begin{equation}
S=\!\int\! d^D\! x {1\over 2g_1}\left(( \nabla {\vec \phi}_1)^2\! + \dots +(
\nabla {\vec \phi}_{N-1})^2\right) +{1\over 2g_2}( \nabla {\vec \phi}_N)^2 .
\label{sigmaN}
\end{equation}
The symmetry breaking pattern is now
${\rm SO}(N)\times {\rm SO}(N-1)\rightarrow {\rm SO}(N-1)_{\rm diag}$. This is
a sigma model which is defined on a space which is homogeneous but not
maximally symmetric. When $g_1 =g_2$, one is dealing with the sigma model
defined by the maximally symmetric space ${\rm SO}(N)\times {\rm SO}(N)/{\rm
SO}(N)_{\rm diag}$, this is the so-called principal chiral model. When
$g_2\rightarrow 0$, it reduces to the principal chiral ${\rm SO}(N-1)$ model,
and when $g_1\rightarrow\infty$ to the ${\rm O}(N)$ vector model. In fact, this
model has been studied some time ago by Friedan\cite{Friedan} who has shown
that it is renormalizable in two dimensions with only two coupling constants.
The corresponding RG flow in $D=2+\epsilon$ is depicted in Fig.~2.

The principal chiral models are known to have a fixed point within the
$D=2+\epsilon$ expansion: these are the points C$_N$ and C$_{N-1}$ in Fig.~2.
The ${\rm O}(N)$ vector model has also a fixed point O$_N$ which is at distance
$\epsilon$ from the upper left-hand corner of the Fig.~2. The novelty is that
we find a fixed point P$_N$ with nontrivial values of the couplings $g_1, g_2$
which is not on the diagonal $g_1 =g_2$. This fixed point has two directions of
instability and thus there are two phase transition lines: one that goes from
P$_N$ to O$_N$ and one that goes from P$_N$ to C$_{N-1}$. This implies that
there is an intermediate phase in addition to the high-temperature paramagnetic
phase and the low-temperature ordered phase. In this new phase, the vector
${\vec \phi}_N$ is ordered because it is in the low-temperature regime of the
${\rm O}(N)$ model but the remaining $N-1$ vectors ${\vec\phi}_1,\dots,
{\vec\phi}_{N-1}$ are still fluctuating in the subspace orthogonal to ${\vec
\phi}_N$.
Due to this partial ordering, we will refer to this phase as ``nematic''. The
transition between the fully ordered phase and the nematic phase is governed by
the fixed point C$_{N-1}$ and the transition between nematic and paramagnetic
phases by the ${\rm O}(N)$ fixed point.
The fixed point C$_N$ which has the highest symmetry governs the critical
behaviour of the direct transition between the paramagnetic phase and the fully
ordered phase.

This phase diagram thus points to a possible explanation of what happens in the
${\rm O}(N)\times O(2)$ sigma model above two dimensions. Here the point
C$_{N-1}$ is replaced by an ${\rm O}(2)$ point which is not seen in the
$D=2+\epsilon$ expansion. This point is located at $g_2 =0$ but $g_1 =O(1)$ and
thus the {\it perturbative} flow of the coupling $g_1$ is that of a
low-temperature phase. The O$(3)$ point is seen in perturbation
theory\cite{Andrey}. The mixed fixed point will also be invisible in
perturbation theory. In fact, in the flow equations of the generalized model,
the limit $N\rightarrow 3$ leads to $g^*_1\rightarrow \infty$ and $g^*_2
=O(\epsilon)$. When $D\rightarrow 2$ the fixed points that were of order
$\epsilon$ collapse to the origin but the ${\rm O}(2)$ fixed point becomes the
KT point at the end of a line of perturbative fixed points.
So the picture we have constructed leads naturally to a phase diagram as Fig.~1
for all ${\rm O}(N)\times {\rm O}(2)$ models.

These $D=2+\epsilon$ expansion results clearly show that the phase diagrams of
the these sigma models is much more complex than previously thought\cite{aza}.
Notably the fixed point P$_N$ is a prominent candidate to interact with the
principal chiral point C$_N$. To deepen our understanding, we now turn to a
mean-field study of the ${\rm O}(3)\times {\rm O}(2)$ sigma model on a
hypercubic lattice. This lattice model is regularized both in the ultraviolet
and in the infrared limit:
this is a sensible way to define the sigma model beyond perturbation theory.
Note that to construct a continuum field theory one has to find a second-order
phase transition which is not guaranteed {\it a priori}. We introduce a field
${\vec \lambda}_i$ conjugate to each vector ${\vec\phi}_i$ in the action
(\ref{sigma3}).
The standard mean-field method then involves a non-Abelian integration over the
${\rm O}(3)$ matrix $({\vec\phi}_1,{\vec\phi}_2,{\vec\phi}_3)$. We start from
the Hamiltonian:
\begin{equation}
H=-\sum_{<x,y>}\! K_i\, {\vec\phi}_i (x)\cdot {\vec\phi}_i (y) \equiv -\sum
K_i\, {\vec\phi}_i \cdot V^{-1}\!\!\cdot {\vec\phi}_i , \label{Hamil}
\end{equation}
where the sum is over nearest-neighbor lattice sites, $V_{xy}$ is the
connectivity matrix and we take only two distinct couplings: $K_1\simeq 1/g_1$
for ${\vec\phi}_1,{\vec\phi}_2$ and $K_2\simeq 1/g_2$ for ${\vec\phi}_3$.
We write the partition function as:
\begin{equation}
Z=\int d{\vec\lambda}_i\, {\rm e}^{-{T\over 4K_i} {\vec\lambda}_i \cdot
V^{-1}\!\cdot {\vec\lambda}_i} \int d\mu ({\vec\phi}_i )\, {\rm
e}^{\mathop{\Sigma}\limits_i
{\vec\lambda}_i \cdot {\vec\phi}_i}.
\label{Zed}
\end{equation}
Here $d\mu$ is the Haar measure on ${\rm SO}(3)$ and T is the temperature. The
mean-field theory is obtained by a saddle-point treatment of the integral over
the auxiliary fields ${\vec \lambda}_i$. We have searched spatially uniform
solutions since the model (\ref{Hamil}) is ferromagnetic. There is a phase
where all the expectation values $\langle {\vec \lambda}_i\rangle $ are zero:
this is the high-temperature fully disordered phase and there is also a phase
where all the $\langle{\vec \lambda}_i\rangle$ are non zero: this is fully
ordered phase that breaks the full rotational invariance. But we also find a
phase where $\langle {\vec \lambda}_3\rangle\neq 0$ but $\langle {\vec
\lambda}_{1,2}\rangle =0$: this is the nematic phase. The corresponding phase
diagram is sketched in Fig.~3.

The most remarkable feature is the appearance of a first-order line that
crosses the diagonal $K_1=K_2$.
Its existence is simple to understand: the non-Abelian integral in
Eq.(\ref{Zed}) has no simple closed form but its expansion in powers of group
invariants built from the matrix
$\Lambda = ({\vec \lambda}_1,{\vec \lambda}_2, {\vec \lambda}_3 )$ is simple.
Since we are dealing with the rotation group there is an odd invariant that
appears in the mean-field potential: $ {\rm Det }[\Lambda ]$. This cubic
contribution gives rise to the first-order transition (thick line in Fig.~3).
This line terminates at tricritical points and is continued by second-order
transition lines:
in particular for $K_2$ small enough there is a second-order transition to the
fully ordered phase.
In the limiting case K$_2=0$, it is not necessary to introduce the auxiliary
field ${\vec \lambda}_3$ in Eq.(\ref{Zed}) and a standard calculation leads
immediately to the Landau theory in Eq.(\ref{LG}). When $0<K_2<K_1$, the field
${\vec \lambda}_3$ is necessarily present but it remains {\it massive} at the
phase transition: At the transition the fields $\vec\lambda_1,\vec\lambda_2$
get a non-zero expectation value, but the $\det[\Lambda]$ term in the potential
acts as a magnetic field and immediately induces also an ordering of
$\vec\lambda_3$. 

The intermediate ordering transitions that occur in the $K_2>K_1$ region are
second-order close to the boundaries. These lines still exist in the unphysical
region below the first-order line: they are then spinodal lines and they
converge right at the diagonal toward the chiral point C$_3$, which is thus
metastable.
As a consequence, we note that the latticized principal ${\rm O}(3)$ chiral
model does not lead to a continuous theory: there is no place where the
correlation length diverges. In fact there is numerical evidence from
Monte-Carlo studies\cite{kunz,diep3} for a first-order transition in the model
(\ref{Hamil}) at $K_1=K_2$ in three dimensions in agreement with the mean-field
prediction. These studies have also obtained marginal evidence for the chiral
universality class exponents of Refs.(\cite{kawa2,bata,diep1}) by simulation of
(\ref{Hamil}) for the value $K_2 =0$.
This Hamiltonian is expected to be in the same universality class as the
STA-type helimagnets. {From} our findings it is however clear that the
proximity
of a tricritical point as seen in mean-field theory may lead to difficulties in
the observation of the true critical behaviour: indeed the first-order line
begins at $K_2 /K_1 =8.5$.

In conclusion, we have shown the existence of a nematic phase with partial spin
ordering in the family of sigma models ${\rm SO}(N)\times {\rm
SO}(N-1)\rightarrow {\rm SO}(N-1)_{\rm diag}$. For $N>3$, all relevant fixed
points are captured by a $D=2+\epsilon$ expansion. In the physical case
SO(3)$\times$SO(2) there is an XY phase transition between the fully ordered
phase and the nematic phase. We obtain a similar picture from mean-field theory
with the appearance of a first-order line that continue the helimagnetic
second-order line between the paramagnetic and the fully ordered phases, and
isolates the principal chiral fixed point C$_3$ with SO(3)$\times$SO(3)
symmetry in the metastability region.
The simplest scenario is that this line appears at
an unknown critical dimension $D_c$ above which the $D=2+\epsilon$ is replaced
by the mean-field picture. In this respect we note that the intersection of the
spinodal lines on the diagonal is suggestive of a collapse of the two fixed
points P$_N$ and C$_N$.  If this critical dimension $D_c$ is between 2 and 3,
there is a natural explanation to the fact that the chiral universality class
has exponents different from O(4).

\acknowledgments
We thank S. Miyashita for an interesting discussion about these topics.

\begin{figure}
\centerline{\epsfxsize=7.0cm\epsfbox{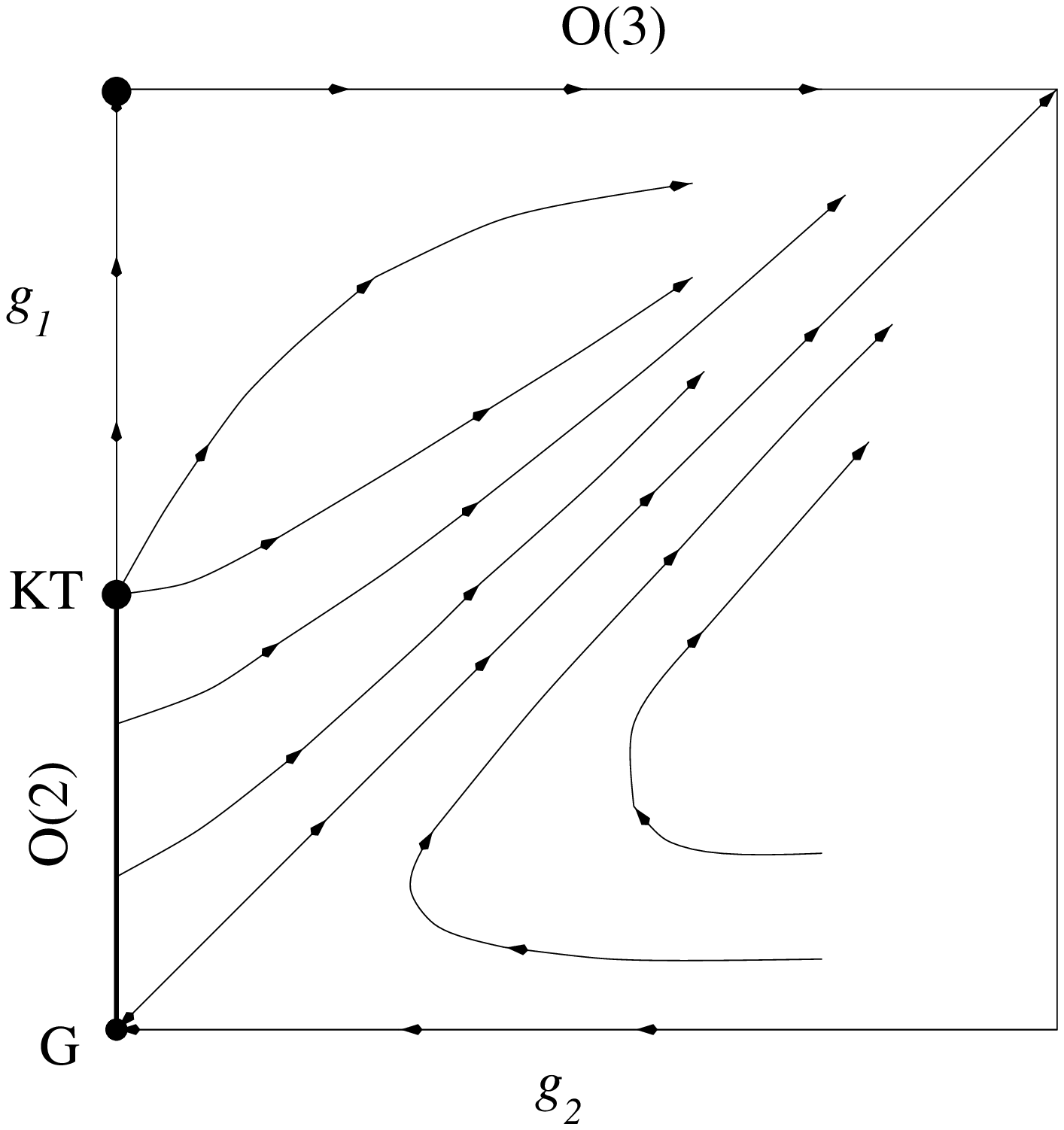}} \medskip
\caption{Global renormalization group flow of the ${\rm O}(3)\times {\rm O}(2)$
sigma model in two dimensions. The perturbative line of fixed points on the
$g_1$ axis ends at a KT fixed point.}
\end{figure}

\begin{figure}
\centerline{\epsfxsize=8.5cm\epsfbox{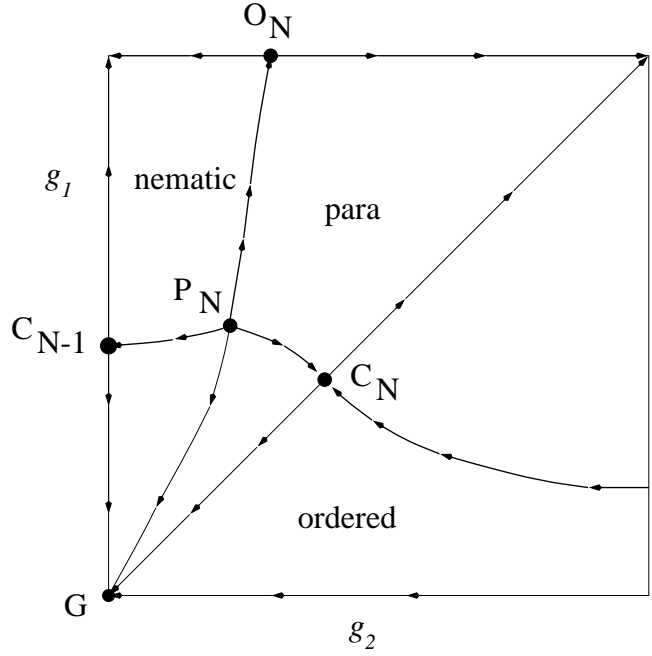}} \medskip
\caption{The RG flow and the corresponding phase diagram of the $O(N)\times
O(N-1)$ sigma model in the neighborhood of two dimensions $D=2+\epsilon$. There
is a high-temperature paramagnetic phase, a fully ordered phase at low
temperature and an intermediate nematic-like phase where the $\phi_N$ vector is
ordered but the other vectors are disordered. The diagonal $g_1 =g_2$ is the
principal chiral model.}
\end{figure}

\begin{figure}
\centerline{\epsfxsize=8.5cm\epsfbox{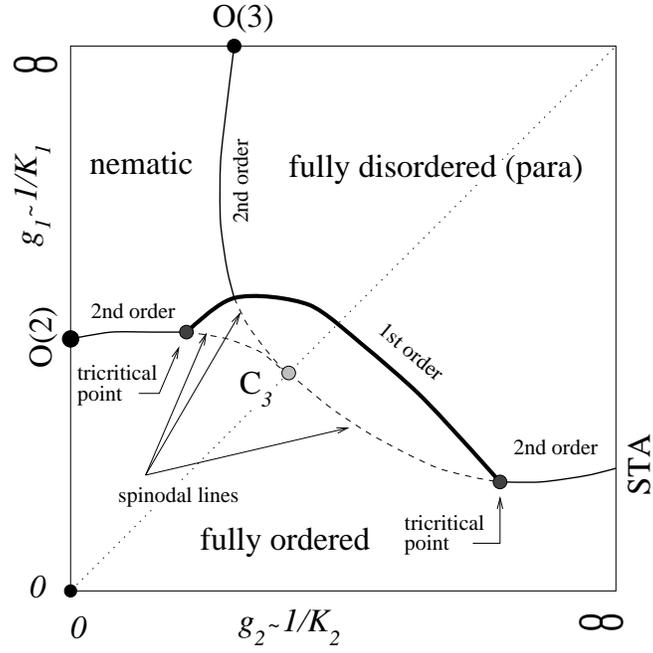}} \medskip
\caption{The non-Abelian mean-field phase diagram of the $O(3)\times O(2)$
sigma model (sketchy).}

\end{figure}
\end{document}